\documentstyle[prl,aps,epsf]{revtex} \def\narrowtext{} \tighten \twocolumn
\input epsf.sty
\begin{document}
\draft
\title{
Direct evidence for an intrinsic square vortex lattice in the overdoped 
high-$T_{c}$ superconductor La$_{1.83}$Sr$_{0.17}$CuO$_{4+\delta}$.}

\author{
	R. Gilardi,$^{1}$
        J. Mesot,$^1$
		A. Drew,$^2$
		U. Divakar,$^2$
		S. L. Lee,$^2$
		E. M. Forgan,$^3$
		O. Zaharko,$^1$
		K. Conder,$^1$
		V. K. Aswal,$^4$
		C. D. Dewhurst,$^5$
		R. Cubitt,$^5$
		N. Momono,$^6$
		and M. Oda.$^6$
		}
\address{f
         $^{1}$ Laboratory for Neutron Scattering, ETH Zurich and PSI 
             Villigen, CH-5232 Villigen PSI, Switzerland\\
		 $^{2}$ School of Physics and Astronomy, University 
		     of St. Andrews, Fife, KY16 9SS, UK\\
		 $^{3}$ School of Physics and Astronomy, University of 
		     Birmingham, Birmingham B15 2TT UK\\
		 $^{4}$ Spallation Neutron Source Division, PSI Villigen, 
		     CH-5232 Villigen PSI, Switzerland\\
		$^{5}$ Institut Laue-Langevin, BP 156, F-38042 Grenoble, France\\
		$^{6}$ Department of Physics, Hokkaido University, 
		     Sapporo 060-0810, Japan\\
			 }
\address{%
\begin{minipage}[t]{6.0in}
\begin{abstract}
We report here the first direct observations of a well ordered vortex lattice in the bulk of a 
La$_{2-x}$Sr$_{x}$CuO$_{4+\delta}$  (La214) single crystal (slightly overdoped, 
x=0.17). Our small angle neutron scattering investigation of 
the mixed phase reveals a crossover from triangular to square coordination 
with increasing magnetic field. The existence of an intrinsic square vortex 
lattice has never been observed in high-temperature superconductors 
(HTSC), and is indicative of the coupling of the vortex lattice to a source of anisotropy, such as 
those provided by a $d$-wave order parameter or the presence of stripes. 
\typeout{polish abstract}
\end{abstract}
\end{minipage}
}

\maketitle
\narrowtext
Apart from the unusual electronic and magnetic behaviour of the cuprate HTSC, experiments reveal a 
tremendously rich variety of mesoscopic phenomena associated with the flux 
vortices in the mixed state\cite{BLATTER94}.
Due to their two dimensional electronic structure, the HTSC are highly anisotropic. The anisotropy 
is characterized by the ratio $\gamma=\lambda_{\perp}/\lambda_{\parallel}$ , 
where $\lambda_{\perp}, \lambda_{\parallel}$ are the superconducting penetration depths for currents flowing
perpendicular and parallel to the two-dimensional CuO$_{2}$ planes. In 
La214 the degree of anisotropy ($\gamma\approx 20$ for x=0.15)
lies between that of the YBa$_{2}$Cu$_{3}$O$_{x}$ (Y123) and 
Bi$_{2}$Sr$_{2}$CaCu$_{2}$O$_{8+x}$ (Bi2212) materials\cite{SASAGAWA00}. The cuprates are also 
extreme type-II superconductors, indicated by the high value of the 
Ginzburg-Landau parameter $\kappa=\lambda/\xi$ , 
where $\xi$ is the superconducting coherence length. In HTSC the combination 
of high transition temperature $T_{c}$, 
high $\gamma$ and high $\kappa$ leads to exotic vortex behaviour, such 
as the phenomenon of vortex lattice (VL) melting\cite{BLATTER94}. 
For a conventional (isotropic) pairing mechanism such anisotropic conduction properties can lead to distortions 
of the vortex lattice as the applied field is tilted towards the CuO$_{2}$ planes, but the 
local coordination remains 
sixfold\cite{THIEMANN89,YETHIRAJ93A,JOHNSON99,YETHIRAJ93B}.

From the magnetic point of view the single-layer (one CuO$_{2}$ sheet per unit cell) La214 compounds appear 
to be rather different from their two-layered Y123 and Bi2212 counterparts. La214 is characterized by the
presence of incommensurate spin excitations\cite{MASON92} located in the vicinity of the antiferromagnetic 
wavevector 
$(\pi,\pi)$ of the undoped parent compound
and at low temperature a spin-gap develops in the spectrum. While such excitations could originate from 
coherence effects in the superconducting 
state\cite{FERMI} it has also been proposed that they could indicate 
the presence of dynamical stripes\cite{TRANQUADA95}. It is still unclear to what extent the presence of 
dynamical stripes 
and/or antiferromagnetic fluctuations affects the vortex state. Recently, inelastic neutron scattering 
experiments in a magnetic field have revealed sub-gap excitations which have been attributed to the existence 
of antiferromagnetic excitations in the vortex core\cite{LAKE01}. It 
was also shown that the opening of the spin-gap is correlated to the 
melting-line of the VL rather than $T_{c}$\cite{LAKE01,MESOT01}. These results indicate that a
subtle interplay may indeed exist between the unusual microscopic and mesoscopic 
properties of the La214 compounds.

Our experiments were performed on the small angle neutron scattering (SANS) 
instruments of both the 
Paul Scherrer Institute (PSI)\cite{KOHL}, Switzerland, and of the Institut 
Laue Langevin (D22), Grenoble, France, with a neutron velocity selector 
of 10\% full width at half maximum, to produce a flux of neutrons with 
a mean wavelength $\lambda_{0}$ between 6 and 15~\AA. 
The incident neutrons were collimated over a distance of 18 m before reaching the sample.
A 96x96 cm$^{2}$ multidetector\cite{SCHLUMPF} at a distance of 18 m was used to detect the scattered neutrons. 

The sample\cite{ODA} was a single crystal cylinder of length 20 mm, 
diameter 6 mm with the c-axis perpendicular to the cylinder.
One of the $\{1,1\}$ in-plane axes, which correspond to the 
directions of the Cu-O bonds of the CuO$_{2}$ planes lay roughly 
along the cylinder axis. 
The crystal had composition La$_{1.83}$Sr$_{0.17}$CuO$_{4}$ ($T_{c}$=37 K).
The width of the superconducting transition, measured by ac-susceptibility, 
was quite small (${\Delta}T_{c}$=1.3 K, as defined by the 10\%-90\%
criterion), indicating the high sample quality.
It was mounted in a cryostat in a magnetic field of 0-1.2~T 
applied parallel to the incident beam. 

For the initial 
measurements, the c-axis was oriented along the field direction.
In a SANS experiment  a conventional VL should give rise to Bragg reflections 
at reciprocal lattice points $q_{hk}$.
The intensity of a single \textit{(h,k)} reflection is given by $I_{hk}\propto 
|F_{hk}|^{2}/q_{hk}$, where $F_{hk}$ is a spatial Fourier component of the field profile in the vortex lattice. 
For the case of a high-$\kappa$ superconductor such as La214 with the field perpendicular to the superconducting planes, 
in the London limit (fields well below the upper critical field) the form factor is 
$F_{hk}= B/(1+{(q_{hk}\lambda_{\parallel}})^{2})$,
where $B$ is the average magnetic induction. This expression may easily be modified to account 
for an anisotropic system\cite{THIEMANN89}. For $B>\mu_{0}H_{c1}$ 
($H_{c1}$ is the lower critical field) the intensity of a first order
reflection reduces to  $I_{10}\propto d_{10}\Phi_{0}^{2}/\lambda_{\parallel}^{4}$
where $q_{10}=2\pi/d_{10}$ and $\Phi_{0}$ is the flux quantum $h/2e$.  In La214, the London penetration 
depth is similar to that measured in Bi2212\cite{SCHNEIDER00}. Although SANS  has been used extensively
to measure VL structure in both the Bi2212\cite{CUBITT93} and Y123 
systems\cite{YETHIRAJ93A,JOHNSON99,YETHIRAJ93B}, to date the La214 family
remains virtually unexplored\cite{VOSTNER98}.

Figure~1 shows the difference of D22-data taken after cooling from 
T=40K$ >T_{c}$ to T=1.5~K at 0.1~T and a background taken at T=40~K.
Figure~2 shows similar PSI-data at 0.8~T with a 
background taken at low temperature in zero-field.
These subtractions are a clean way to remove the large background signal arising from
small angle defect scattering from the crystal. Due to the large mosaic 
of the vortex lattice, several spots
could be observed simultaneously. We could furthermore rotate the cryomagnet and sample about the
vertical and horizontal axes, allowing us to collect the integrated magnetic intensity as we rocked 
the Bragg spots through the Ewald sphere. Figure~3a shows the 
tangential average of the neutron signal,
as a function of the modulus of the wavevector \textbf{q}. As expected, the position 
(in reciprocal space) of the peak maximum changes with the field, thus clearly establishing the 
VL origin of the neutron signal.

We now return to the lowest-field data (B=0.1~T) shown in 
Figure~1. When the c-axis lies along the field direction and the 
field is kept constant during cooling, we observe a large number of spots distributed 
around a ring (Fig.~1a, high-resolution mode: $\lambda_{0}=14.5~\AA$). Since the value of the wave 
vector obtained from the tangential average is close to the one expected for a triangular 
lattice (see Fig.~3b), the pattern most likely consists of
a superposition of diffraction from various domain orientations of 
triangular coordination.
In order to confirm this triangular coordination, it is necessary to 
reduce the high degree of 
degeneracy apparent in Figure~1a. This was achieved by rotating the 
c-axis (about the vertical axis)
away from the field direction by a sufficiently large angle $\Theta$. 
Further improvement of the mosaic spread of the VL was obtained by 
oscillating the field around its mean value while cooling.
The results of these procedures are shown in Fig.~1b ($\Theta = 
10^{\circ}$, $\lambda_{0}=8~\AA$) where a triangular 
pattern can be observed.
Our measurements shown in Fig.~1 are very reminiscent
of what has been seen in untwinned Y123\cite{JOHNSON99}  crystals. In that 
system, a small amount of residual 
twin planes appears to control the orientation of the VL,
giving rise to a high level of degeneracy. As the c-axis is rotated away 
from the field direction, the triangular symmetry 
becomes evident. 

As the field is increased from 0.1~T to 0.8~T a completely different  pattern emerges, with all of
the magnetic scattering concentrated in four intense spots appearing along the $\{1,1\}$ directions, 
forming a perfect square (Figure~2a).
To establish the significance of this fourfold diffraction pattern at higher field we must first 
take account of the influence of the twin planes. While the alignment of such patterns has previously
been attributed to the influence of an anisotropic vortex core in 
Y123\cite{KEIMER94}, more recent measurements on
untwinned Y123 crystals\cite{JOHNSON99} have confirmed earlier 
suggestions\cite{YETHIRAJ93A,FORGAN95} that these results in twinned
crystals are actually explained in terms of  strong alignment of the triangular 
VL mainly due to 
twin-plane pinning. Thus in Y123 when the field is roughly perpendicular 
to the ab planes, an intrinsic 
triangular coordination may give rise to a predominantly fourfold diffraction pattern. This conclusion 
is also inferred in twinned Y123 by rotating the field away from the c-axis so that the influence of 
the twins is severely reduced, resulting in the recovery of a triangular 
coordination\cite{JOHNSON99}. In Y123 for small
tilt angles $\Theta<\Theta_{c}\approx 5^{o}$ a fraction of vortices remains pinned to the c-axis, since 
the vortices bend in order 
to lie for part of their length within the planar 
defects\cite{BLATTER94,YETHIRAJ93A}. For larger angles this is no longer 
energetically favourable and the vortices lie along the direction of the applied field\cite{YETHIRAJ93B}. 
Estimates of this critical angle $\Theta_{c}$ for La214 
would not suggest a significantly different behaviour from Y123, 
given the slightly increased values for $\lambda$ and $\gamma$, 
although the precise value is difficult to predict\cite{BLATTER94}
and falls with increasing field. 
Our measurements at low field (see Fig.1) indicate that for our 
La214 compound $\Theta_{c}$ is lower than $10^{\circ}$. 
Hence, for data taken at high field, we can be sure that $\Theta = 30^{o}$
exceeds the critical angle. We see in Figure~2b  that a four-spot 
pattern is retained at this angle. 
Thus it is extremely
unlikely that this pattern arises in La214 from pinning distortions due to the presence of twin
boundaries, and these data represent the first such observations of an 
intrinsically square VL 
in a cuprate HTSC. For completeness we note that for $\Theta = 30^{o}$ there exists a slight distortion of 
the square pattern which is in accord with expectations for fields tilted 
away from the $c$-axis. Due to the approximately uniaxial superconducting 
anisotropy, the tilting causes an increase of 
the penetration depth of the vortex currents along one direction, distorting the vortex shape
and hence the vortex lattice. In the London approach for anisotropic uniaxial 
superconductors\cite{CAMPBELL88} 
a rhomboid is expected. The expected values for the 
ratio $\epsilon=a/b$ between the sides ($a$ and $b$) of the rhomboid and 
for the internal angle $\beta$ are 
$\epsilon$=0.88, $\beta=86.4^{o}$. A two dimensional fit of the neutron data yields $\epsilon=0.90(3)$ 
and $\beta=87.3^{o}(1.7^{o})$,
which agree well with the theoretical values. Furthermore, the scattered intensity changes
roughly as $cos^{2}(\Theta)$ as expected from the changes in the penetration depth. This indicates that 
all of the vortices lie along the field direction and not along the 
$c$-axis, that is, they are not pinned within the twin planes.
Further contributory evidence for a square coordination comes from the positions of the Bragg spots 
in reciprocal space. The relationship between the magnetic field $B$ and 
the magnitude $q$ of the
wavevector  depends on a structure-dependent quantity $\sigma=(2\pi/q)^{2}B/\Phi_{0}$,
where $\sigma$ is equal to $\sqrt{3}/2$ or 1 for triangular and square lattices, respectively. 
At high magnetic field, the experimental values of $\sigma$ are as expected 
for a square lattice, 
while at low field they are consistent with that of a triangular 
lattice (see Figure~3b). An alternative way to 
quantify the triangular to square transition consists in monitoring the intensity ratio of sectors 
($\pm 15^{\circ})$ containing the $\{1,0\}$ and $\{1,1\}$-type directions. As shown in 
Figure~3b this intensity decreases steadily from 1 at 
B=0.1 T to about 0.2 at B=0.4 T and then remains constant. 

The experimental evidence thus indicates that an intrinsic square vortex lattice exists at
fields 0.5~T and larger. We now discuss the possible origins of this fourfold coordination. 
Isotropic London or Ginzburg-Landau (GL) theories predict a triangular 
VL in high-$\kappa$ systems 
such as the cuprates.  This is also true for modified theories which take into account the
anisotropy of  layered materials such as the HTSC. VL of square coordination have however been observed
in non-cuprate superconductors and may occur for a number of reasons. In 
low-$\kappa$ materials non-local 
electrodynamic effects are expected to be more significant due to stronger interactions between 
the cores of neighbouring vortices. This may give rise to a VL of square coordination, although
at low magnetic induction (large vortex separation) or close to $T_{c}$ (reduced influence of non-local 
effects) the VL tends to a triangular coordination. In combination with anisotropic electronic structure,
this explains the recent observations of triangular to square transitions in the nickel boron-carbide 
superconductors\cite{PAUL98}. A square VL can also arise in extended GL theories which allow for the existence 
of more than one order parameter, which may account for the robustness 
of the square VL in the 
candidate $p$-wave system Sr$_{2}$RuO$_{4}$\cite{RISEMANN98}. A fourfold symmetry 
could also result from the anisotropic ($d$-wave) 
nature of the superconducting gap\cite{BERLINSKY95,XU96,SHIRAISHI99,ICHIOKA99} 
via the increasing importance of the anisotropic vortex cores 
at high field.  The onset of such effects might be expected to occur at fields a factor of order ten 
lower than in Y123 ($T_{c}^{max}=93K$), due to the smaller gap in La214 
($T_{c}^{max}=38.5K$). The experimental search 
for such a transition has been complicated in Y123 by the influence of twin-plane pinning discussed above. 
To date  there are no reports on either  twinned or untwinned Y123 for such an intrinsic change of 
coordination. The only evidence for a structural change as a function of field is a reorientation of
the triangular lattice\cite{JOHNSON99} starting at fields of around 3~T. In Bi2212 the vortex lattice cannot be 
observed using SANS at high fields, due to the disorder induced by point 
pinning\cite{CUBITT93}. It is interesting
to notice that the early VL-calculations based on d-wave gap 
functions\cite{BERLINSKY95,XU96} predict the square 
lattice to be oriented along the Cu-O-Cu bonds ($\{1,1\}$ direction in LSCO), which is exactly what our
experimental data reveal. However, more recent 
VL-calculations\cite{SHIRAISHI99,ICHIOKA99} favor a square lattice tilted 
by $45^{o}$ from the Cu-O-Cu bonds. Our results provide an important 
stimulus to resolve these theoretical discrepancies.
In principle other sources of anisotropy, such as those involving the presence of dynamical 
stripes\cite{KIVELSON98} or charge-density waves\cite{MOOK99}, could lead 
to the formation of a square vortex 
lattice. However, the plausibility of this explanation is as yet unclear,  since our measurements 
were performed in the overdoped regime of the HTSC where a homogeneous Fermi liquid description 
of the electronic properties of these materials is currently believed to be adequate. 

In conclusion we have used SANS to make the first clear observation of the vortex lattice 
in LSCO and obtained strong evidence for a field-induced crossover from a triangular to a square 
lattice. This may reflect the increasing importance of the anisotropic vortex cores in this 
$d$-wave superconductor, or coupling to other sources of anisotropy such 
as those provided by charge/stripe fluctuations.
It remains a challenge to corroborate detailed SANS measurements with other microscopic data 
in order to explain the origin of this exotic vortex behaviour.

We thank T. Konter, J. Kohlbrecher, A. Bollhalder, P. Rasmussen, and N. Schlumpf for 
the technical assistance.
This work was supported by the Swiss National Science Foundation, 
the Engineering and Physical Sciences Research Council of U.K. 
and the Ministry of Education and Science of Japan.

\begin{figure}
\epsfxsize=3.0in
\epsfbox{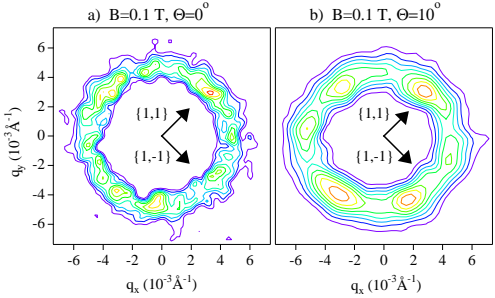}
\vspace{0.5cm}
\caption{
SANS diffraction patterns taken at T=1.5~K, after field cooling 
from 40~K in B=0.1~T. A background taken at T=40~K has been subtracted. 
One of the $\{1,1\}$ in-plane axes is aligned at 45 degrees to the horizontal axis
in the plane perpendicular to the field.
In a) the c-axis lies along the field direction and the field was kept constant during 
cooling, high-resolution mode: $\lambda_{0}=14.5~\AA$. In b) the c-axis has been rotated 10$^{\circ}$ (about the vertical axis) 
away from the field direction, and while cooling the 
field was oscillated about its mean value, $\lambda_{0}=8~\AA$.
}
\end{figure}

\begin{figure}
\epsfxsize=3.0in
\epsfbox{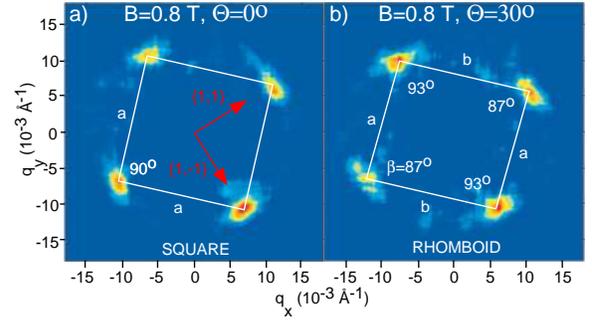}
\vspace{0.5cm}
\caption{
SANS diffraction patterns obtained by subtracting the zero-field 
data at 5~K from the data taken at 5~K for a field B=0.8~T applied parallel 
to the beam, $\lambda_{0}=8~\AA$.  One of the $\{1,1\}$ in-plane axes is aligned at 32 degrees to the horizontal axis
in the plane perpendicular to the field. The c-axis of the sample is either a) parallel to the field or
b) rotated 30 degrees away from it. The result of the two-dimensional fits 
of the diffraction patterns are shown in a) by the white square and in b) by the white 
rhomboid which illustrates the distortion induced by the rotation of the c-axis away 
from direction of the applied magnetic field.}
\end{figure}

\begin{figure}
\epsfxsize=3.2in
\epsfbox{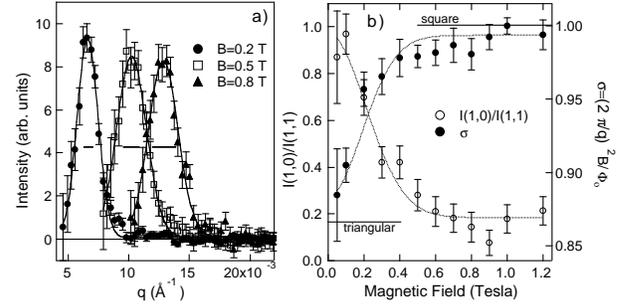}
\vspace{0.5cm}
\caption{
a) Tangential average of the neutron signal 
for B=0.2 T (circles), 0.5 T (squares) and 0.8 T (triangles) at T=5~K. 
The intensities have been rescaled to allow a common vertical axis. 
The shift in $q$ illustrates the expected dependence of the vortex lattice spacing on increasing 
magnetic field. The instrumental resolution (FWHM) is given by the horizontal 
segments.
b) The black circles show the field dependence of 
$\sigma=(2\pi/q)^{2}B/\Phi_{0}$  at T=1.5~K. 
The horizontal lines represent the expected values of 
$\sigma$ for triangular and square lattices. The 
open circles show the intensity ratio of sectors
containing the $\{1,0\}$ and $\{1,1\}$ directions.}
\end{figure}

\end{document}